\pdfoutput=1
\documentclass{article}

\usepackage[preprint]{neurips_2026}
\workshoptitle{Representations for the Physical Sciences Workshop}

\usepackage[utf8]{inputenc}
\usepackage[T1]{fontenc}
\usepackage[hidelinks]{hyperref}
\usepackage{url}
\usepackage{booktabs}
\usepackage{amsfonts}
\usepackage{amsmath}
\usepackage{amssymb}
\usepackage{nicefrac}
\usepackage{microtype}
\usepackage{graphicx}
\usepackage{subcaption}
\usepackage{tikz}

\graphicspath{{figures/}}

\newcommand{\unifoil}{UniFoil}
\newcommand{\Cl}{C_\ell}
\newcommand{\Cm}{C_m}
\newcommand{\Ma}{\mathrm{Ma}}
\newcommand{\appref}[1]{Appendix~\ref{#1}}

\title{How Does Distribution Shift Shape Pretraining Gains in Neural PDE Surrogates?} %Attributing Transfer Gains to Shift Components\\
\author{%
  Pochinapeddi Sai Bhargav \\
  Rensselaer Polytechnic Institute \\
  Troy, NY 12180 \\
  \texttt{pochis@rpi.edu} \\
  \And
  Nithin Somasekharan \\
  Rensselaer Polytechnic Institute \\
  Troy, NY 12180 \\
  \texttt{somasn@rpi.edu} \\
  \AND
  Rohit Sunil Kanchi \\
  University of Tennessee \\
  Knoxville, TN 37996 \\
  \texttt{rkanchi@vols.utk.edu} \\
  \And
  Sicheng He \\
  University of Tennessee \\
  Knoxville, TN 37996 \\
  \texttt{sicheng@utk.edu} \\
  \AND
  Shaowu Pan\thanks{Corresponding author.} \\
  Rensselaer Polytechnic Institute \\
  Troy, NY 12180 \\
  \texttt{pans2@rpi.edu} \\
}
\begin{document}

\maketitle

\begin{abstract}
Pretraining a neural PDE surrogate can reduce the amount of new CFD data
needed when geometry or modeled physics changes. However, it remains unclear
how different components of distribution shift affect this benefit. We
pretrain a surrogate on $254{,}909$ RANS solutions from one airfoil family
and fine-tune it on a new family under two target settings with matched
freestream ranges: the same Spalart--Allmaras (SA) modeling and SA with
added $e^N$ transition modeling. At $N=1000$, the pretrained
model matches the accuracy of a model trained from scratch on $3.25\times$
as many samples for the same-SA target, but $2.58\times$ as many for the
transition-modeled target. By $N=5000$, this ordering reverses
($1.56\times$ versus $1.86\times$). At $N=1000$, sampling more distinct airfoils lowers error on both targets, but only for the same-SA target is the gain increase larger than the observed draw-to-draw variation ($3.3\times$ to $4.0\times$). These results show that pretraining value depends jointly on target-data budget, target-data coverage, and whether source and target differ in modeled physics.

\end{abstract}

% OWNS: the problem, the attribution gap, why airfoil aerodynamics is the right
% instrument, why existing benchmarks cannot close the gap, contributions.
% DOES NOT OWN: corpus counts beyond the pretraining size, split methodology,
% any result, any mechanism.
\section{Introduction}
\label{sec:intro}

Neural PDE surrogates can be expensive to build because each training sample may require a costly numerical
solution. This cost recurs whenever the geometry, operating regime, or modeled physics changes. 
PDE foundation models~\citep{subramanian2023towards,Herde2024Poseidon} promise to turn that recurring cost into a one-time investment: pretrain once on a large PDE corpus, then fine-tune on a few samples from each new regime, even when its distribution differs from the pretraining data. On canonical PDE benchmarks, this approach reduces downstream
sample requirements by one to two orders of magnitude. 
Pretraining datasets that span multiple PDE
systems are now expressly assembled for such reuse \citep{McCabe2024MPP,Hao2024DPOT}.

Existing PDE foundation models report pretraining gains across PDE systems, geometries, and operating conditions. However, evaluations often simplify distribution shift to a single PDE system parameter, such as Reynolds number, leaving the pretrained regime~\citep{subramanian2023towards,simshift2025}. 
% Existing PDE foundation models are mostly pretrained on corpora spanning many distinct PDE systems, where governing physics, geometry and operating conditions change all together.  existing benchmarks grade the departure along a single axis, pushing one PDE
% parameter out of range \citep{subramanian2023towards,simshift2025}. 
% 
Realistic distribution shifts in PDE problems often combine multiple distinct \emph{shift components}, such as geometry, governing equations, and system parameters. A practitioner deciding whether to reuse a pretrained representation therefore needs to understand how each component affects transfer-learning gains from a pretrained neural PDE surrogate, which we call \emph{pretraining gains}. 
% , not only in degree, and a saving measured under such a compound shift credits no
% particular component --- yet a practitioner deciding whether to reuse a pretrained representation needs to know which source--target differences erode that benefit. 
We analyze these effects by varying one shift component at a time. 
% We therefore probe the question where it can be answered cleanly: within a single PDE system whose shift components vary one at a time. What generalizes is the protocol, not the corpus.

In this work, we study neural PDE surrogates for two-dimensional aerodynamics using the \unifoil{}~\citep{Kanchi2025UniFoil} datasets, which contain solutions of the steady Reynolds-Averaged Navier--Stokes (RANS) equations across airfoil geometries, modeled-physics configurations, and operating conditions. 
Unlike the earlier airfoil benchmark dataset~\citep{Bonnet2022AirfRANS}, \unifoil{} contains samples from two geometrically distinct airfoil families over matched freestream ranges that extend into the transonic regime, together with an additional subset that models laminar-to-turbulent transition. We use three subsets: (1) $254{,}909$ samples from fully turbulent (FT) airfoil geometries with the SA turbulence model; (2) $36{,}412$ samples from natural-laminar-flow (NLF) airfoil geometries with the SA turbulence model; and (3) $34{,}585$ samples from natural-laminar-flow airfoil geometries with the SA turbulence model and $e^N$ transition model. 
This design isolates the effect of geometry on pretraining gains by 
pretraining on the first geometry family and fine-tuning on the second.
Fine-tuning on the second geometry family with added laminar-to-turbulent transition modeling then measures the cumulative effect of geometry and modeled-physics shifts on pretraining gains. 
% fine-tuning instead on the same NLF shape distribution under an added laminar-to-turbulent transition model measures how the transfer gain changes when the modeled
% physics changes as well. 
% In this work, we intentionally fix the freestream conditions identical across all three datasets. 
% of 
% Labels come from a solver, so the corpus can hold the freestream ranges identical across all three subsets; $254{,}909$ pretraining samples make the question meaningful at scale.

\paragraph{Contributions.} 
% (1) We construct a controlled distributional comparison: one target changes the airfoil-family distribution at fixed closure, while the second adds transition modeling, with matched freestream ranges, geometry-disjoint splits and a from-scratch control at every budget.
(1) We formulate a controlled comparison missing from existing PDE-pretraining evaluations~\citep{yang2026toward}. Existing evaluations either change several source--target factors at once or vary only a scalar system parameter. We instead compare an NLF geometry-family target under the source SA modeling configuration with an NLF target that additionally introduces $e^N$ transition modeling, while matching the freestream ranges.
% 
% (2) At $N=1000$ the sample-efficiency gain is $3.3\times$ for the same-closure geometry shift and $2.6\times$ once transition modeling is added. The gap is positive under all nine pairings of the observed data draws, and its bootstrap interval for one fresh replication excludes zero.
(2) We identify an interaction between target sample budget and modeled-physics shift. Pretraining gains decrease with target sample budget on both fine-tuning tasks (see Figure 4 in Appendix), but their ordering reverses. At $N=1000$,
the same-SA geometry target has the larger sample-efficiency gain
($3.25\times$ versus $2.58\times$), whereas by $N=5000$ the
transition-modeled target has the larger gain ($1.86\times$ versus
$1.56\times$). Thus, modeled-physics mismatch does not impose a fixed
penalty on pretraining gain; its observed effect depends on the amount of available target data.
% 
% (3) Most of the measured benefit remains accessible without changing a pretrained backbone: a conv-adapter recovers $83\%$ of full fine-tuning's improvement, whereas the same adapter on a
% never-pretrained backbone falls short of training from scratch.
% (4) At a fixed training budget of $N=1000$, allocating samples across more distinct airfoils
% lowers error on both targets and raises the gain from $3.3\times$ to $4.0\times$ under the
% same-closure geometry shift. Under the transition-modeled shift, the gain does not rise resolvably,
% consistent with the missing source information being modeled physics rather than geometric
% coverage.
(3) We find that target-data allocation interacts with the composition of distribution shift. At a fixed budget of $N=1000$, spreading samples across more
distinct airfoils lowers error on both targets, but produces a resolvable increase in
the sample-efficiency gain only under the same-SA geometry shift, from $3.3\times$ to $4.0\times$. 
However, on the transition-modeled target, the gain
changes slightly from $2.6\times$ to $2.7\times$ with overlapping per-draw ranges.
Thus, sampling across more distinct airfoils amplifies the value of pretraining only when it addresses variability not accompanied by a change in modeled physics.
% \begin{itemize}
% % \itemsep1pt\parsep0pt\topsep1pt\partopsep0pt
% \item We construct a controlled distributional comparison: one target changes the airfoil-family distribution at fixed closure, while the second adds transition modeling, with matched freestream ranges, geometry-disjoint splits and a from-scratch control at every budget.
% \item At $N=1000$ the sample-efficiency gain is $3.3\times$ for the same-closure geometry shift and $2.6\times$ once transition modeling is added. The gap is positive under all nine pairings of the observed data draws, and its bootstrap interval for one fresh replication excludes zero.
% \item Most of the measured benefit remains accessible without changing a pretrained backbone: a conv-adapter recovers $83\%$ of full fine-tuning's improvement, whereas the same adapter on a
% never-pretrained backbone falls short of training from scratch.
% \item At a fixed training budget of $N=1000$, allocating samples across more distinct airfoils
% lowers error on both targets and raises the gain from $3.3\times$ to $4.0\times$ under the
% same-closure geometry shift. Under the transition-modeled shift, the gain does not rise resolvably,
% consistent with the missing source information being modeled physics rather than geometric
% coverage.
% \end{itemize}

\section{Experimental setup}
\label{sec:setup}

\paragraph{Pretraining corpus and prediction task.}
\unifoil{} provides steady two-dimensional RANS samples over two airfoil families from distinct
design distributions: fully-turbulent (FT) sections and natural-laminar-flow (NLF) sections, the
latter shaped to hold the boundary layer laminar over much of the chord
\citep{Kanchi2025UniFoil}. We use three subsets
totaling $465{,}286$ samples; the FT subset supplies $363{,}841$ of these, of which the $254{,}909$
training samples form the pretraining corpus (Table~\ref{tab:shift}). The surrogate maps a discretized geometry and its three freestream
scalars --- drawn independently and uniformly over $\Ma_\infty\!\in\![0.10,0.85]$,
$\alpha\!\in\![-2^\circ,6^\circ]$, $Re\!\in\![10^6,10^7]$ --- to the four flow fields
$(u,v,\Ma,C_p)$ on the body-fitted grid; lift and quarter-chord moment follow by integrating the surface pressure. These ranges are identical across the three subsets, so no part of any measured gap can come from
extrapolation in the PDE parameters ($\Ma_\infty,\alpha,Re$) --- the axis along which existing
benchmarks induce shift --- leaving geometry and modeled physics as the components under study.

\paragraph{Geometry-disjoint evaluation.}
We split by airfoil geometry, not by sample: every sample for an airfoil geometry goes entirely into either train, validation
or test. An i.i.d.\ split would divide one airfoil's conditions between training and test, so the
model would meet every test geometry at some other condition and the evaluation would measure
interpolation across the envelope rather than generalization to unseen shape. Every reported number
therefore measures generalization to unseen geometry: a model fine-tuned on $N=1000$ target samples is
evaluated on over $10{,}000$ held-out-geometry samples per target.

\begin{table}[t]
\centering\footnotesize
\caption{Pretraining corpus and the two fine-tuning targets. Freestream ranges and the splitting
rule are identical throughout, so the two shift components accumulate one at a time. A closure is
the modeling that closes the RANS equations: SA is the Spalart--Allmaras turbulence model;
SA $+$ $e^N$ adds transition prediction. Counts are train/val/test; symbols in \appref{app:notation},
geometry counts in \appref{app:geomcount}.}
\label{tab:shift}
\vspace{2pt}
\setlength{\tabcolsep}{6pt}
\begin{tabular}{llllc}
\toprule
Dataset & Geometry & Closure / transition & Samples (train/val/test) & Shift vs.\ pretraining \\
\midrule
$\mathcal{D}_\mathrm{FT}$        & FT airfoils  & SA                & 254{,}909 / 36{,}514 / 72{,}418 & --- (source) \\
$\mathcal{D}_\text{NLF-Turb}$    & NLF airfoils & SA                & 36{,}412 / 5{,}318 / 10{,}143  & geometry \\
$\mathcal{D}_\text{NLF-Trans}$   & NLF airfoils & SA $+$ $e^N$      & 34{,}585 / 4{,}876 / 10{,}111  & geometry $+$ transition modeling \\
\bottomrule
\end{tabular}
\end{table}

\paragraph{The decomposed distribution shift: NLF-Turb and NLF-Trans.}
The two targets step away from the pretraining distribution one component at a time
(Table~\ref{tab:shift}). NLF-Turb draws NLF geometries under the same one-equation closure as the
source; NLF-Trans draws from the same NLF shape distribution over the same ranges but adds an $e^N$
transition model, so laminar--turbulent transition is modeled rather than assuming fully turbulent flow. Both targets draw from the same shape database over the
same ranges, so their fine-tuning curves differ by what the transition model adds. The body-fitted grid is $84\times292$ for the source
against $84\times304$ for both targets, so the circumferential resolution changes alongside the geometry,
but it is common to the two targets.

\paragraph{Architecture selection: convolutional U-Net.}
Five architectures spanning the families in current use for flow-field prediction were compared on
the pretraining task alone at a matched parameter budget; the convolutional U-Net \citep{Ronneberger2015UNet,Thuerey2020} was the most
accurate on every quantity and among the least expensive to train (\appref{app:arch}). We held this
architecture fixed across all transfer experiments reported here (\appref{app:hparams}). The selection used no target
data, so the choice of backbone is independent of the transfer outcome; what remains open is whether the
ordering of shift components holds across architecture families.

\paragraph{The efficiency metric: from-scratch vs. pretrained.}
Every fine-tuned model is paired with a from-scratch control: the identical network trained on the same
target samples from random initialization. Reading the two error-versus-budget curves horizontally,
at fixed accuracy, gives the \emph{sample-efficiency gain}: the factor by which pretraining shrinks the fine-tuning training set needed to
reach a given accuracy, with the budget spent on many conditions over few airfoils (\emph{depth})
unless stated otherwise.

\paragraph{Zero-shot accuracy: the shift is real.}
Before any fine-tuning, the pretrained model's lift error on the two targets is $0.88$ and $0.94$ of
each target's mean $|\Cl|$ --- barely better than predicting zero lift
(\appref{app:zeroshot}). Absence from training does not explain it: held-out source-family
geometries are equally absent and predicted accurately, so the degradation is distributional, not a
matter of unseen shape. Output scale does not explain the degradation either
(\appref{app:peft}). Fig.~\ref{fig:fieldzs} (\appref{app:fields}) shows one held-out geometry zero-shot
and after fine-tuning.

\section{Results}
\label{sec:results}

\begin{figure}[t]
\centering
\begin{subfigure}[t]{0.485\linewidth}
\centering
\includegraphics{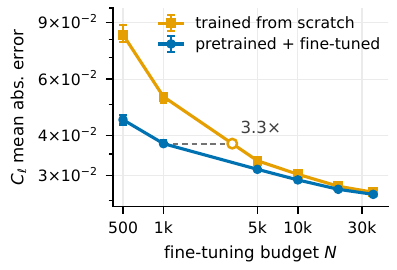}
\caption{Geometry shift (NLF-Turb)}
\label{fig:de_turb}
\end{subfigure}\hfill
\begin{subfigure}[t]{0.485\linewidth}
\centering
\includegraphics{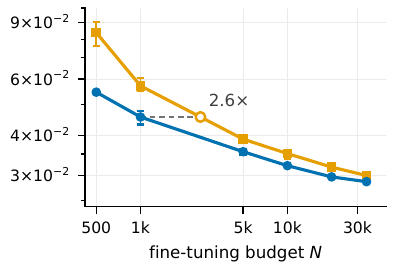}
\caption{Geometry $+$ transition-modeling shift (NLF-Trans)}
\label{fig:de_trans}
\end{subfigure}
\caption{Held-out lift error against fine-tuning budget $N$, log--log: pretrained and fine-tuned
(circles) versus the from-scratch control on identical samples (squares). Bars span three
independent data draws, each a full retraining rather than evaluation noise; full-data points are single
runs. The dashed read-off at matched error is the sample-efficiency gain at $N=1000$:
$3.3\times$ under the geometry shift~(a), $2.6\times$ when transition modeling is added~(b).}
\label{fig:de_both}
\end{figure}

\textbf{(Q1)} How many fine-tuning samples does pretraining save, and does the saving depend on
the shifted component? \textbf{(Q2)} What does the pretrained representation supply?
\textbf{(Q3)} How does the saving depend on how the budget is spent?

\paragraph{(Q1) Sample-efficiency gains by shift component.}
\looseness=-1
The gain depends on the target distribution. At $N=1000$ a pretrained model reaches a lift accuracy
the from-scratch control does not attain until roughly $3{,}250$ training samples, a gain of
$3.3\times$ under the same-closure geometry shift (Fig.~\ref{fig:de_turb}); the corresponding
measurement on the transition-modeled target gives $2.6\times$ (Fig.~\ref{fig:de_trans}), about a
fifth less. Fig.~\ref{fig:fieldfs} (\appref{app:fields}) shows the two models at equal budget on one
held-out geometry. We report lift ($\Cl\;\text{MAE}$) throughout because it is defined identically
across all three corpora, is a central aerodynamic quantity, and gives the most conservative of the
four measured gains on both targets (\appref{app:metrics}). The ordering persists across the
observed data draws: their gains span $3.16$--$3.38$ and $2.35$--$2.81$, so all nine cross-target
pairings are positive. A bootstrap over data draws and held-out airfoils places the gap at
$+0.68\times$, with a $95\%$ interval of $[0.03,1.25]$ for one fresh replication
($B=10{,}000$; \appref{app:uncertainty}).

\paragraph{Decay of both gains with budget.}
\looseness=-1
Both gains shrink beyond $N=1000$ and their ordering reverses by $N=5000$: the geometry-shift
gain falls from $3.3\times$ to $1.4\times$ between $N=1000$ and $N=10{,}000$, whereas the transition-modeled
gain falls only from $2.6\times$ to $1.7\times$ (\appref{app:decay}). In the single full-data runs, the control closes the NLF-Turb gap to within $0.0004$, whereas a
$0.0013$ gap remains on NLF-Trans; the observed advantage therefore persists to a larger budget on
the harder target.

\paragraph{(Q2) What the pretrained representation supplies.}
\looseness=-1
Freezing every pretrained weight and adding a parallel convolutional adapter ($5\%$ more parameters)
reaches $0.0412$, recovering $83\%$ of full fine-tuning's improvement over the from-scratch control.
The same adapter on an identically shaped backbone that was never pretrained reaches $0.1117$,
worse than the $0.0537$ obtained by training from scratch, so adapter capacity alone cannot explain
the result. Most of the measured benefit therefore remains accessible through convolutional
adaptation of fixed pretrained features, while full fine-tuning supplies the remainder (geometry
shift, depth allocation, $N=1000$, one data draw; \appref{app:peft}).

\begin{table}[t]
\centering\small
\caption{Depth versus breadth sampling at $N=1000$: pretrained held-out lift error and the resulting
sample-efficiency gain for both targets. Means over three data draws; parenthesized ranges are the
per-draw spread. \textbf{Bold} marks the largest gain.}
\label{tab:protocol}
\vspace{2pt}
\setlength{\tabcolsep}{10pt}
\begin{tabular}{lcccc}
\toprule
 & \multicolumn{2}{c}{depth} & \multicolumn{2}{c}{breadth} \\
\cmidrule(lr){2-3}\cmidrule(lr){4-5}
Target & $\Cl$ MAE & gain & $\Cl$ MAE & gain \\
\midrule
NLF-Turb  & 0.0377 & 3.3$\times$ (3.16--3.38) & 0.0342 & \textbf{4.0$\times$} (3.58--4.36) \\
NLF-Trans & 0.0457 & 2.6$\times$ (2.35--2.81) & 0.0416 & 2.7$\times$ (2.28--3.37) \\
\bottomrule
\end{tabular}
\end{table}

\paragraph{(Q3) Geometric diversity versus parameter coverage.}
\looseness=-1
The budget can instead favor geometric diversity --- at $N=1000$, roughly one operating
condition per airfoil (\emph{breadth}) --- rather than covering the
PDE parameters ($\Ma_\infty,\alpha,Re$) more densely over far fewer airfoils, as depth does. Breadth gives lower held-out error for both targets in every data draw
(Table~\ref{tab:protocol}), so it is the better allocation at this budget. Its effect on the
pretraining premium differs by target. Under the same-closure geometry shift, the gain rises from
$3.3\times$ to $4.0\times$ with disjoint per-draw ranges, consistent with the target set using its
limited budget to add geometric novelty while inheriting parameter coverage from pretraining. Under
the transition-modeled shift, the gain changes only from $2.6\times$ to $2.7\times$ with overlapping
per-draw ranges. Because neither allocation supplies transition-modeled source information, this
unresolved change is consistent with missing modeled physics limiting the transfer premium. At
a matched geometry count and larger budgets, additional operating conditions remain valuable
(\appref{app:geomcount}).

% OWNS: the summary claim, the scoped limitation, one forward statement.
% DOES NOT OWN: any number not already reported, any new argument. Framing is
% BENEFIT (samples saved), never "cost" -- the paper measures what pretraining
% buys, and the conclusion must not switch currency.
\section{Conclusion}
\label{sec:conclusion}

\looseness=-1

Our results suggest that distribution shift does not impose a simple fixed penalty on the pretraining gains of neural PDE surrogates. Instead, the value of pretraining for downstream fine-tuning depends jointly on the amount and composition of target data and on what the source and target share. 
Pretraining should therefore be viewed as a source--target compatibility problem rather than a universal guarantee of data efficiency. Predicting this compatibility across PDE tasks is an important step toward reusable scientific foundation models.

\bibliographystyle{plainnat}
\bibliography{refs}

\appendix

\section{Notation}
\label{app:notation}

Every abbreviation and symbol used in the body is described here.

\vspace{2pt}
\noindent\textit{Abbreviations}
\vspace{2pt}

\noindent\begin{tabular}{@{}l @{\quad=\quad} p{0.85\linewidth}@{}}
RANS & Reynolds-averaged Navier--Stokes equations \\
closure & the modeling that closes the RANS equations; the paper treats it as one component of the distribution shift \\
SA & Spalart--Allmaras, a one-equation turbulence model, run here without a transition model, so the boundary layer is turbulent from essentially the leading edge \\
$e^N$ & a transition model added on top of SA that predicts where the boundary layer turns from laminar to turbulent \\
FT & Fully turbulent airfoils (conventional) sections --- the pretraining (source) family. It refers to the design intent, not to the closure \\
NLF & natural-laminar-flow airfoils --- the fine-tuning (target) family \\
MAE & mean absolute error \\
rel-$L_2$ & relative $L_2$ error: the $L_2$ norm of the prediction residual divided by that of the reference field \\
\end{tabular}

\vspace{6pt}
\noindent\textit{Symbols}
\vspace{2pt}

\noindent\begin{tabular}{@{}l @{\quad=\quad} p{0.78\linewidth}@{}}
$\Ma_\infty,\ \alpha,\ Re$ & freestream Mach number, angle of attack and chord Reynolds number: the three scalars fixing the operating condition, and the model's non-geometric \emph{inputs} \\
$u,\ v$ & the two in-plane velocity components (model \emph{outputs}) \\
$\Ma$ & local Mach-number field (a model \emph{output}, distinct from the freestream $\Ma_\infty$) \\
$C_p$ & pressure coefficient (model \emph{output}); its value on the airfoil surface gives the loads \\
$\Cl,\ \Cm$ & lift and quarter-chord pitching-moment coefficients, obtained by integrating the surface $C_p$ \\
$c$ & airfoil chord, the straight-line length from leading to trailing edge \\
$N$ & number of labeled target samples used for fine-tuning \\
$B$ & number of bootstrap replicates \\
$\mathcal{D}_\mathrm{FT}$ & pretraining corpus: FT airfoils under the SA closure \\
$\mathcal{D}_{\text{NLF-Turb}}$ & first target: NLF airfoils under the same SA closure --- a change of geometry distribution alone \\
$\mathcal{D}_{\text{NLF-Trans}}$ & second target: NLF airfoils under SA $+$ $e^N$ --- the geometry change plus a change of closure \\
\end{tabular}

\section{Geometry coverage and budget allocation}
\label{app:geomcount}

This appendix supplies the geometry-count analysis supporting the allocation contrast of
Table~\ref{tab:protocol}, discussed as (Q3) in Sec.~\ref{sec:results}. The contrast is not reproduced
here: Table~\ref{tab:protocol} already carries all four values.

The two allocations reach comparable geometry counts at very different budgets, which separates
geometric variety from sample count. Breadth at $N=1000$ covers $851$ airfoils; depth at
$N=10{,}000$ covers $777$. Almost the same number of geometries, ten times the samples --- and the
latter is $15\%$ more accurate on the geometry-shift target ($0.0290$ against $0.0342$) and $23\%$
on the transition-modeled one ($0.0322$ against $0.0416$). Additional conditions on already-seen
geometries therefore continue to help substantially once geometry count is held fixed, so the
allocation result of Sec.~\ref{sec:results} is not a claim that geometry count alone determines
accuracy.

The comparison is reported at $N=1000$ because the contrast is sharpest there and must close at
larger budgets: the fine-tuning pools hold only $2{,}818$ and $2{,}517$ distinct airfoils, and
breadth has drawn on more than $95\%$ of them by $N=10{,}000$ and more than $99\%$ by
$N=20{,}000$, so the geometric variety it can still add is largely spent.

\paragraph{Coverage of the two target corpora.}
Sampling the same shape database twice does not return the same sections: across all splits
NLF-Turb covers $4{,}012$ distinct airfoils and NLF-Trans $3{,}615$, with $1{,}833$ in common. The difference is in shapes rather than cases: the corpus computes $2{,}179$ NLF shapes only under SA and $1{,}782$ only with the transition model, so NLF-Trans draws on $7$--$11\%$ fewer geometries while carrying slightly more operating conditions per geometry ($13.7$ against $12.9$ in training, $13.7$ against $12.7$ in test). The two test sets are within $0.3\%$ in size ($10{,}111$ against $10{,}143$). The corpus computes only the NLF family under both modeling configurations, so the shift components are measured cumulatively: the effect of added transition modeling is its increment given the geometry change, never transition modeling alone.

The breadth gains of Table~\ref{tab:protocol} are read against from-scratch controls trained on
breadth-allocated manifests at every budget and data draw, so the two columns of that table compare
like with like.

Restricting the held-out set to the $377$ airfoils the two targets have in common leaves the
geometry-shift gain essentially unchanged ($3.19\times$ against $3.25\times$ on the full test set)
and raises the transition-modeled one from $2.58\times$ to $2.64\times$. The per-draw ranges stay disjoint on the matched
subset ($3.13$--$3.30$ against $2.27$--$2.93$), so every cross-target pairing remains positive on
matched geometries. Split
assignment is a property of the airfoil rather than of the corpus: an airfoil appearing in both
corpora falls in the same split in both, so no target's test geometries appear in the other's
fine-tuning data.

\section{Architecture comparison at matched budget}
\label{app:arch}

The backbone was fixed before any transfer experiment, on the pretraining task alone, by a
matched-budget comparison of five architectures spanning the families in current use for flow-field
prediction (Table~\ref{tab:arch}): a convolutional U-Net \citep{Ronneberger2015UNet,Thuerey2020},
Transolver \citep{wu2024transolver}, a ViT \citep{dosovitskiy2021vit}, an FNO
\citep{li2021fourier} and CViT \citep{Wang2025CViT}. The U-Net is the most accurate candidate on every quantity while
being among the least expensive to train; the closest competitor costs roughly $25\times$ the
GPU-hours (Table~\ref{tab:archcost}), which places a comparison across several sizes or model initializations beyond
the compute budget.

Its accuracy is set by the architecture rather than by capacity: an $8\times$ parameter increase
leaves every metric unchanged, and the ordering is not even monotonic (Table~\ref{tab:scaling}). The
S tier is therefore the efficient operating point and the one used throughout. This is a selection
for a controlled study, not a claim that the U-Net is universally superior --- each architecture was
trained under its own authors' recommended configuration on one corpus at one resolution, and a
different corpus, grid or budget could reorder them.

\begin{table}[h]
\centering\small
\caption{Candidate networks at a matched parameter budget ($\sim$22\,M), on held-out-geometry test
samples in physical (denormalized) units. Lower is better; \textbf{bold} marks the lowest value in
each row. All models are trained till convergence (converged by 100 epochs) and the lowest validation error model is used for test set inference.}
\label{tab:arch}
\vspace{2pt}
\setlength{\tabcolsep}{7pt}
\begin{tabular}{lccccc}
\toprule
Held-out-geometry metric & U-Net-S & Transolver-S & ViT-S & FNO-S & CViT-S \\
\midrule
$u$ rel-$L_2$ [\%]   & \textbf{3.15}   & 3.53   & 4.35   & 4.04   & 4.48   \\
$v$ rel-$L_2$ [\%]   & \textbf{5.95}   & 6.25   & 8.55   & 8.25   & 9.45   \\
$\Ma$ rel-$L_2$ [\%] & \textbf{3.03}   & 3.39   & 4.18   & 3.90   & 4.33   \\
$C_p$ surface MAE    & \textbf{0.0236} & 0.0265 & 0.0370 & 0.0359 & 0.0432 \\
$\Cl$ MAE     & \textbf{0.0266} & 0.0283 & 0.0313 & 0.0315 & 0.0328 \\
$\Cm$ MAE     & \textbf{0.0065} & 0.0069 & 0.0076 & 0.0079 & 0.0079 \\
\midrule
Training cost [GPU-h] & $\sim$29 & $\sim$727 & $\sim$69 & \textbf{$\sim$20} & $\sim$56 \\
\bottomrule
\end{tabular}
\end{table}

\begin{table}[h]
\centering\small
\caption{U-Net width scaling on the held-out-geometry test set ($72{,}418$ samples). An $8\times$ parameter range leaves every metric unchanged. \textbf{Bold} marks the lowest
value in each row.}
\label{tab:scaling}
\vspace{2pt}
\setlength{\tabcolsep}{9pt}
\begin{tabular}{lcccc}
\toprule
Quantity & T (11\,M) & S (22\,M) & B (35\,M) & L (90\,M) \\
\midrule
Field rel-$L_2$ mean [\%] & \textbf{1.127} & 1.136 & 1.141 & 1.131 \\
$\Ma$ rel-$L_2$ [\%]      & \textbf{3.004} & 3.030 & 3.040 & 3.019 \\
$C_p$ surface MAE         & 0.0242 & \textbf{0.0236} & 0.0241 & 0.0236 \\
$\Cl$ MAE          & 0.0271 & \textbf{0.0266} & 0.0270 & 0.0269 \\
$\Cm$ MAE          & 0.0066 & \textbf{0.0065} & 0.0066 & 0.0066 \\
\bottomrule
\end{tabular}
\end{table}

\begin{table}[h]
\centering\small
\caption{Measured training cost at the S tier for four of the five candidates; ViT-S is priced in Table~\ref{tab:arch}. GPU-hours are wall
time multiplied by the device count of the row's hardware.}
\label{tab:archcost}
\vspace{2pt}
\setlength{\tabcolsep}{10pt}
\begin{tabular}{lcccc}
\toprule
Model (S tier) & Hardware & Wall time & GPU-hours & vs.\ U-Net \\
\midrule
U-Net-S      & 4$\times$A100  & $\sim$7.2\,h & $\sim$29  & 1$\times$ \\
FNO-S        & 4$\times$H100  & $\sim$5.1\,h & $\sim$20  & 0.7$\times$ \\
CViT-S       & 8$\times$GPU   & $\sim$7.0\,h & $\sim$56  & $\sim$1.9$\times$ \\
Transolver-S & 16$\times$A100 & 45.4\,h      & $\sim$727 & $\sim$25$\times$ \\
\bottomrule
\end{tabular}
\end{table}

\section{Training configuration}
\label{app:hparams}

Table~\ref{tab:hparams} gives the configuration used for pretraining, fine-tuning and training from
scratch. Apart from the training horizon, the network, optimizer, schedule form and loss are shared
throughout. Fine-tuned models train for 150 epochs, whereas from-scratch controls receive 300
epochs because they converge more slowly; all runs were verified to have converged from their
training and validation histories. The evaluated checkpoint is the one with the lowest error on the
target validation split, which is fixed across every budget and shared by the fine-tuned model and
its from-scratch control.

\begin{table}[h]
\centering\small
\caption{U-Net-S configuration for the transfer study. The network is identical across FT
pretraining, from-scratch training and target fine-tuning; only the initialization and the
trainable-parameter set change.}
\label{tab:hparams}
\vspace{2pt}
\begin{tabular}{@{}ll@{}}
\toprule
Setting & Value \\
\midrule
Backbone & U-Net (conv encoder--decoder, GroupNorm, GELU) \\
Base width / channel mult.\ / levels & 56 / $(1,2,4,8,16)$ / 5 \\
Parameters & 22.2\,M \\
Input / output channels & $[x,y,\Ma,\alpha,Re]$ / $[u,v,\Ma,C_p]$ \\
Grid & $84\times292$ (FT) / $84\times304$ (NLF) O-grid \\
Precision / loss & fp32 / MSE in min--max-normalized space \\
Optimizer & AdamW, lr $10^{-3}$, weight decay $0.01$ \\
Schedule & 5-epoch warmup, cosine to $10^{-8}$, grad-clip $1.0$ \\
Batch size & 8 \\
Checkpoint selection & lowest error on the target validation split of Table~\ref{tab:shift} \\
Pretraining epochs & 100 \\
Fine-tuning / from-scratch epochs & 150 / 300 \\
Convergence check & all training and validation histories verified converged \\
Data draws & 3 (depth), 3 (breadth); single at full data \\
Fine-tuning initializations & 1 throughout, 3 in \appref{app:seeds} \\
Pretrained checkpoint & shared by every transfer experiment \\
Hardware & NVIDIA A100 / H100 \\
\bottomrule
\end{tabular}
\end{table}

\section{Source-family generalization and zero-shot transfer}
\label{app:zeroshot}

Table~\ref{tab:ftgen} establishes that the pretrained model generalizes within its own family:
held-out test accuracy matches held-out validation accuracy on every quantity, so the network is
predicting unfamiliar airfoils rather than reproducing memorized ones. Table~\ref{tab:zeroshot}
gives its accuracy on the two targets before any fine-tuning. Against the $0.0266$ of
Table~\ref{tab:ftgen} these are degradations of $19\times$ and $12\times$; the normalized figures
quoted in Sec.~\ref{sec:setup} divide them instead by each target's mean $|\Cl|$.

The zero-shot magnitudes are not a difficulty ordering. NLF-Turb's larger absolute lift error
reflects the larger lift magnitudes of its held-out samples (mean $|\Cl|$ $0.56$ against $0.35$);
normalized by them the ordering reverses ($0.88$ against $0.94$), leaving NLF-Trans marginally the
worse of the two, as the pretrained errors at $N=1000$ also indicate on all four metrics
(Table~\ref{tab:metricsboth}).

\begin{table}[h]
\centering\small
\caption{Held-out FT geometry generalization before any NLF fine-tuning, on $72{,}418$
unseen-geometry test samples. Mean \% (norm) is field error in the normalized training
space over $u$, $v$, $\Ma$, $C_p$; Mean \% (phys) is the physical equivalent over $u$, $v$, $\Ma$.
$\Cl$ and $\Cm$ are pressure-integrated. Lower is better.}
\label{tab:ftgen}
\vspace{2pt}
\setlength{\tabcolsep}{4pt}
\begin{tabular}{lrrrrrr}
\toprule
Split & Mean \% (norm) & Mean \% (phys) & $C_p$ surface MAE & $\Cl$ MAE & $\Cm$ MAE \\
\midrule
Validation & 1.131 & 4.032 & 0.0235 & 0.0266 & 0.0064 \\
Test       & 1.136 & 4.046 & 0.0236 & 0.0266 & 0.0065 \\
\bottomrule
\end{tabular}
\end{table}

\begin{table}[h]
\centering\small
\caption{Zero-shot evaluation of the FT-pretrained model on held-out target geometries, before any
fine-tuning. Lower is better.}
\label{tab:zeroshot}
\vspace{2pt}
\setlength{\tabcolsep}{8pt}
\begin{tabular}{llrrr}
\toprule
Target & Shift & $C_p$ surface MAE & $\Cl$ MAE & $\Cm$ MAE \\
\midrule
NLF-Turb  & NLF shapes, SA closure       & 0.2782 & 0.4951 & 0.0913 \\
NLF-Trans & NLF shapes, transition model & 0.2440 & 0.3300 & 0.0933 \\
\bottomrule
\end{tabular}
\end{table}

\section{Fine-tuning depth as a probe of the shift}
\label{app:peft}

How much of the network must change to correct the shift is ordinarily a compute question; here it
doubles as a probe of the shift's nature. If the two families differed only in the \emph{scale} of
features the network already computes, re-fitting output statistics would close the gap. It does
not: normalization-and-head fine-tuning, updating $0.05\%$ of weights, is no better than
discarding the pretrained weights and training from scratch --- worse in three of the four
target-allocation cells and indistinguishable in the fourth ($0.0602$ against $0.0605$ on NLF-Trans
under depth; Table~\ref{tab:peftfull}). Output recalibration alone is therefore insufficient; useful transfer requires either changing the pretrained features or learning transformations over them.

Most of the accuracy, though, is recoverable without touching them. A parallel convolutional adapter
\citep{chen2024convadapter}, which freezes every pretrained weight and adds modules amounting to a further
$5\%$ of parameters, reaches $0.0412$ against full fine-tuning's $0.0387$ on the geometry-shift
target at $N=1000$ under depth allocation. Full fine-tuning supplies the remaining improvement not recovered by the frozen-backbone adapter. Under breadth allocation that remainder
grows with the shift, its margin over the conv-adapter widening from $0.0024$ on the geometry-shift
target to $0.0043$ when transition modeling is added; under depth it narrows instead ($0.0025$ to $0.0009$).
The two allocations disagree, so we draw no trend from this comparison.

\paragraph{A frozen backbone that was never pretrained.}
The adapter result could in principle reflect the adapter's own capacity rather than the
representation it recombines. Repeating both frozen-backbone strategies on a randomly initialized
backbone of the same architecture --- same initialization scheme as the from-scratch control, frozen
before training, with the adapter hyperparameters, schedule and checkpoint rule unchanged ---
separates the two. On NLF-Turb the conv-adapter reaches $0.1117$ against $0.0412$ pretrained, and
normalization-and-head $0.5526$ against $0.0566$; on NLF-Trans, $0.1150$ against $0.0487$ and
$0.3497$ against $0.0602$. Both random-frozen rungs are worse than training from scratch
($0.0537$ and $0.0605$), and the ordering is the same on moment, surface pressure and field error.
The adapter's benefit therefore depends on information learned during pretraining rather than on adapter capacity alone.

\begin{table}[h]
\centering\small
\caption{Held-out-geometry $\Cl$ error at $N=1000$ by fine-tuning strategy, target and sampling
allocation, at a single data draw. Lower is better; \textbf{bold} marks the lowest value in each
column. These per-strategy values differ from the multi-draw anchors of Table~\ref{tab:headline} by at
most $6\%$.}
\label{tab:peftfull}
\vspace{2pt}
\setlength{\tabcolsep}{8pt}
\begin{tabular}{lcccc}
\toprule
& \multicolumn{2}{c}{NLF-Turb} & \multicolumn{2}{c}{NLF-Trans} \\
\cmidrule(lr){2-3}\cmidrule(lr){4-5}
Strategy (trainable) & breadth & depth & breadth & depth \\
\midrule
zero-shot (0\%)                 & \multicolumn{2}{c}{0.4951} & \multicolumn{2}{c}{0.3300} \\
from-scratch (100\%$^{*}$)      & 0.0453 & 0.0537 & 0.0528 & 0.0605 \\
normalization-and-head (0.05\%) & 0.0478 & 0.0566 & 0.0602 & 0.0602 \\
parallel conv-adapter ($+5\%$)  & 0.0353 & 0.0412 & 0.0464 & 0.0487 \\
decoder-only (35\%)             & 0.0358 & 0.0399 & 0.0479 & 0.0493 \\
full fine-tuning (100\%)        & \textbf{0.0329} & \textbf{0.0387} & \textbf{0.0421} & \textbf{0.0478} \\
\bottomrule
\end{tabular}
\vspace{1pt}

{\footnotesize\raggedright $^{*}$All parameters trainable, but from random initialization: the
from-scratch row is the control, not a fine-tuning strategy.\par}
\end{table}

\section{What the shift looks like in the field}
\label{app:fields}

The error figures elsewhere in this paper are integrated quantities. This appendix shows one
held-out NLF-Trans geometry directly, so the shift and the effect of fine-tuning can be seen rather than inferred.
The case is representative rather than extreme: its zero-shot lift error is $0.242$ against a
target-wide mean of $0.3300$ (Table~\ref{tab:zeroshot}).

\begin{figure}[h]
\centering
\includegraphics{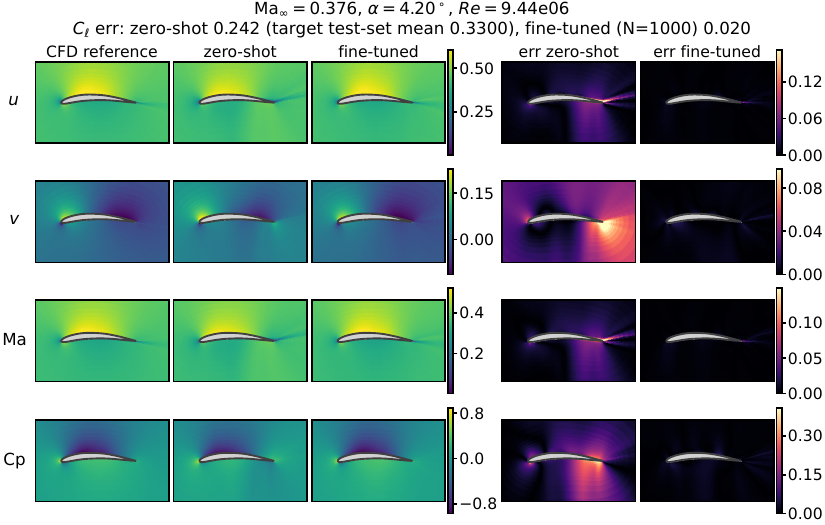}
\caption{Zero-shot against fine-tuned, one held-out NLF-Trans geometry at $\Ma_\infty=0.376$,
$\alpha=4.20^\circ$, $Re=9.44\times10^{6}$. Rows are the four predicted fields; the left block gives
the CFD reference, the pretrained model before any fine-tuning, and the same model after fine-tuning
on $N=1000$ target samples, on a shared colour scale per row. The right block gives the
corresponding absolute errors, each row on its own scale.}
\label{fig:fieldzs}
\end{figure}

Fig.~\ref{fig:fieldzs} separates the zero-shot residual from the fine-tuned one. Before any target data, the pretrained
model reproduces the broad structure of every field but misplaces the suction-side detail, and the
error concentrates on the upper surface and the trailing-edge wake --- exactly where the transition
model changes the boundary layer. After fine-tuning on $N=1000$ target samples the residual is
diffuse and an order of magnitude smaller, and lift error falls from $0.242$ to $0.020$.

\begin{figure}[h]
\centering
\includegraphics{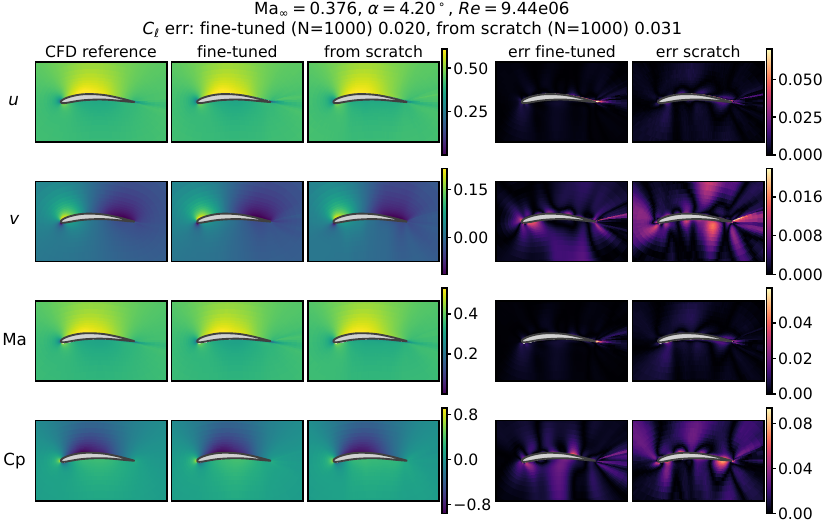}
\caption{Fine-tuned against the from-scratch control at the same budget, same geometry and
operating point as Fig.~\ref{fig:fieldzs}. Both models see $N=1000$ target samples; only the
initialization differs. Layout follows Fig.~\ref{fig:fieldzs}.}
\label{fig:fieldfs}
\end{figure}

Fig.~\ref{fig:fieldfs} makes the comparison the paper actually measures: the same budget spent
from a pretrained initialization against a random one. Both models capture the fields, but the
from-scratch control retains visible structure in the $v$ and $C_p$ residuals where the fine-tuned
model does not, and its lift error is $0.031$ against $0.020$ --- the per-case counterpart of the
budget-averaged gap in Fig.~\ref{fig:de_both}.

\section{Consistency of the ordering across metrics}
\label{app:metrics}

Table~\ref{tab:metricsboth} gives the accuracy underlying the four-metric claim of
Sec.~\ref{sec:results}. At $N=1000$ the from-scratch-to-pretrained error ratio is larger for the
geometry shift than for the geometry-plus-transition-modeling shift on every metric: $1.40$ against $1.25$ on
lift, $1.51$ against $1.39$ on moment, $1.65$ against $1.43$ on surface pressure, and $1.43$ against
$1.23$ on field error. No quantity has saturated at the largest budget measured --- field error
still falls from $2.80\%$ to $1.94\%$ on NLF-Trans between $N=1000$ and full data --- so a growing
budget continues to buy accuracy. What it diminishes is the return on pretraining: the $\Delta$
column falls from $0.0152$ to $0.0004$ on lift for NLF-Turb, and analogously on every metric and on
both targets.

The gain itself varies with the quantity it is read on, which is why the body reports the most
conservative. At $N=1000$ it is $3.3\times$ on lift, $3.7\times$ on moment, $4.5\times$ on the flow
field and $5.3\times$ on surface pressure for NLF-Turb, against $2.6\times$, $2.9\times$, $3.1\times$
and $3.7\times$ for NLF-Trans: lift is the lowest of the four on both targets, and the geometry shift
holds the larger gain on all four. Lift is also the quantity the engineering deliverable rests on,
so reporting on it is deliberately conservative, yet it remains the correct metric for quantifying the benefit of pretraining under the distribution shift.

\begin{table}[h]
\centering\small
\caption{Accuracy at $N=1000$ and at full data, pretrained against from-scratch, both targets,
depth allocation, three-data-draw mean. Lower is better; $\Delta$ is scratch $-$ pretrained, the
benefit pretraining confers at that budget.}
\label{tab:metricsboth}
\vspace{2pt}
\setlength{\tabcolsep}{9pt}
\begin{tabular}{lcccccc}
\toprule
& \multicolumn{3}{c}{$N=1000$} & \multicolumn{3}{c}{full data} \\
\cmidrule(lr){2-4}\cmidrule(lr){5-7}
Metric & pretrained & scratch & $\Delta$ & pretrained & scratch & $\Delta$ \\
\midrule
\multicolumn{7}{@{}l}{\emph{NLF-Turb} (geometry shift; full $N=36{,}412$)}\\
$\Cl$ MAE          & 0.0377 & 0.0529 & 0.0152 & 0.0262 & 0.0266 & 0.0004 \\
$\Cm$ MAE          & 0.0093 & 0.0140 & 0.0047 & 0.0065 & 0.0066 & 0.0001 \\
$C_p$ surface MAE  & 0.0406 & 0.0668 & 0.0262 & 0.0288 & 0.0297 & 0.0009 \\
field rel-$L_2$ \% & 1.6014 & 2.2900 & 0.6886 & 1.1550 & 1.1709 & 0.0159 \\
\midrule
\multicolumn{7}{@{}l}{\emph{NLF-Trans} (geometry $+$ transition-modeling shift; full $N=34{,}585$)}\\
$\Cl$ MAE          & 0.0457 & 0.0571 & 0.0114 & 0.0287 & 0.0300 & 0.0013 \\
$\Cm$ MAE          & 0.0098 & 0.0136 & 0.0038 & 0.0058 & 0.0061 & 0.0003 \\
$C_p$ surface MAE  & 0.0443 & 0.0633 & 0.0190 & 0.0288 & 0.0306 & 0.0018 \\
field rel-$L_2$ \% & 2.7972 & 3.4447 & 0.6475 & 1.9414 & 2.0140 & 0.0726 \\
\bottomrule
\end{tabular}
\end{table}

\section{Uncertainty quantification for the gap between two targets}
\label{app:uncertainty}

This appendix documents the procedure behind (Q1)'s robustness result, in the same three-source
order and the same
vocabulary the body uses. Two definitions, given once. A \emph{case} is a single held-out sample:
one geometry at one operating point, scored by absolute lift error. A comparison is \emph{paired}
when the pretrained model and its from-scratch control are scored on the same resampled cases, so
difficulty common to the two cancels; pairing is available within a target and never between them,
since NLF-Turb and NLF-Trans have disjoint test sets of $10{,}143$ and $10{,}111$ cases.

The quantity is a difference of gains, $\Delta = g_\text{NLF-Turb} - g_\text{NLF-Trans}$ at
$N=1000$, and each $g$ is a curve-crossing statistic: the gain at budget $N$ is the from-scratch
budget at which the control matches the pretrained model's error at $N$, divided by $N$; at
$N=1000$, a matching budget of $3{,}251$ gives $3.25\times$. The crossing is therefore part of the
estimator and is re-fitted inside every replicate rather than once outside the loop.

The from-scratch curve is measured at $N\in\{500,\,1000,\,5000,\,10{,}000,\,20{,}000\}$ and at full
data. It is made monotone by running minimum, and the crossing is located by linear interpolation in
$\log N$ against $\log e$ between the two measured budgets that bracket it; at $N=1000$ that bracket
is $[1000,5000]$ on both targets. Gains are formed per data draw and then averaged, and a draw whose
curve does not reach the pretrained error within the measured range is censored rather than clipped.

\paragraph{Which cases landed in the test set.}
Each target is resampled independently. Within a target, cases are drawn with replacement, the same
resample is applied to the pretrained model and its control at every budget, the log--log
interpolation is re-fitted, and the crossing is read off. Replicates whose crossing falls outside
the measured budget range are recorded as censored rather than clipped; the censoring rate is zero
here, so the interval is not an artifact of truncation. Taken alone --- that is, holding the data
draw fixed --- this marginal analysis gives, at the case level, a $95\%$ percentile interval of
$[0.53,1.10]$ over $B=10{,}000$ replicates.

\paragraph{Which target samples were drawn.}
The three data draws are independent retrainings on different samples of the target pool. Taken
alone they give per-draw gains of $3.16$--$3.38$ (NLF-Turb) and $2.35$--$2.81$ (NLF-Trans): ranges
that do not overlap, so the ordering holds draw-for-draw without any resampling model.
Equivalently, all nine pairings of one NLF-Turb draw with one NLF-Trans draw give a positive gap,
the smallest $+0.36\times$. This is the strongest statement the design supports and it assumes nothing about
the sampling distribution.

\paragraph{Both together.}
The two analyses above are marginal: the first conditions on one data draw, the second on one test
set. Resampling the draw jointly with the cases propagates both, and this is what the reported
interval does --- each replicate first draws a data draw uniformly from the three, then resamples
that draw's test set. The unit resampled is the airfoil, not the case: a
geometry contributes about thirteen operating conditions to the test split, and treating those as
independent is the same assumption the geometry-disjoint split exists to avoid. Each replicate
therefore draws airfoils with replacement and takes all of a drawn airfoil's cases. Over $B=10{,}000$ replicates this gives $\Delta=+0.68\times$; the $95\%$ bootstrap interval for
one fresh replication is $[0.03,1.25]$ over $797$ and $740$ held-out airfoils. Resampling cases instead would give $[0.17,1.15]$,
narrower by $20\%$ --- the cost of the independence assumption, not a real gain in precision. The
interval is for the gap one fresh replication of the study would produce, not for the precision of
the three-draw mean, and the widening over the case-only analysis measures how much of the
uncertainty lives in the training draw rather than the evaluation set.

\paragraph{Sensitivity to the read-off rule.}
The two-point rule locates the crossing inside the bracket $[1000,5000]$, which contains no
measured budget. Refitting the same measured points under two other interpolants moves the point
values and leaves the ordering: the gap is positive under all three.

\begin{table}[h]
\centering\small
\caption{Sensitivity of the sample-efficiency gain to the read-off rule at $N=1000$, on the
three-draw mean curves. No retraining: all three interpolants are applied to the same measured
budgets. The two-point row reads $3.26\times$ against Table~\ref{tab:headline}'s $3.25\times$
because it is computed on the mean curves rather than by averaging the per-draw gains.}
\label{tab:estsens}
\vspace{2pt}
\setlength{\tabcolsep}{10pt}
\begin{tabular}{lccc}
\toprule
read-off rule & NLF-Turb & NLF-Trans & gap \\
\midrule
two-point log--log (used throughout) & 3.26$\times$ & 2.56$\times$ & $+0.70$ \\
monotone cubic (PCHIP), all budgets  & 2.81$\times$ & 2.21$\times$ & $+0.60$ \\
global power law                     & 4.83$\times$ & 3.60$\times$ & $+1.23$ \\
\bottomrule
\end{tabular}
\end{table}

\section{Robustness to initialization}
\label{app:seeds}

The transfer study varies the data draw to isolate sampling variance while holding the model
initialization fixed. Tables~\ref{tab:seedpre} and~\ref{tab:seedscratch} close that gap for the
configuration the $3.3\times$ and $2.6\times$ gains come from --- full fine-tuning, depth
allocation, $N=1000$ ---
by holding the data draw fixed and varying only the model initialization over three values.

Two things follow. First, model-initialization variation is small relative to the effects reported in the
body: the pretrained $\Cl$ range spans $0.0360$--$0.0394$ and $0.0436$--$0.0478$, well inside the
pretrained-versus-scratch gaps of Table~\ref{tab:headline}. Second, and more informative, the
from-scratch range is markedly wider on the same metric --- $4.1\times$ on NLF-Turb and $2.1\times$
on NLF-Trans. The two arms carry different randomness, and that is the point: the pretrained
checkpoint is shared by every transfer experiment, so the fine-tuned model has no weight
initialization to vary and the three runs differ only in the fine-tuning seed, while the control
varies the whole network. Pretraining therefore removes one source of run-to-run variation outright
and leaves a narrower spread in what remains. Both ranges are absolute, and the control's mean error
is the higher, so the ratio should not be read as a scale-free sensitivity.

Read together, the two tables also show the reported configuration to be conservative:
initialization~1 gives the pretrained model its median (NLF-Turb) and its highest (NLF-Trans) lift
error, while giving the from-scratch control its lowest on both targets. Both differences narrow
the measured gap, so the reported gains understate the initialization-averaged benefit.

\begin{table}[h]
\centering\small
\caption{Model-initialization sensitivity of the \emph{pretrained and fine-tuned} model, depth
allocation, $N=1000$, full fine-tuning, data draw held fixed. Initialization~1 is the configuration
the body reports.}
\label{tab:seedpre}
\vspace{2pt}
\setlength{\tabcolsep}{6pt}
\begin{tabular}{lcccc}
\toprule
& $\Cl$ MAE & $\Cm$ MAE & $C_p$ surface MAE & Field rel-$L_2$ [\%] \\
\midrule
\multicolumn{5}{@{}l}{\textit{NLF-Turb}} \\
init 1   & 0.0387 & 0.0095 & 0.0414 & 1.626 \\
init 2   & 0.0394 & 0.0101 & 0.0420 & 1.597 \\
init 3   & 0.0360 & 0.0087 & 0.0418 & 1.629 \\
mean     & 0.0380 & 0.0094 & 0.0417 & 1.617 \\
min--max & 0.0360--0.0394 & 0.0087--0.0101 & 0.0414--0.0420 & 1.597--1.629 \\
\midrule
\multicolumn{5}{@{}l}{\textit{NLF-Trans}} \\
init 1   & 0.0478 & 0.0103 & 0.0454 & 2.807 \\
init 2   & 0.0440 & 0.0091 & 0.0437 & 2.724 \\
init 3   & 0.0436 & 0.0091 & 0.0440 & 2.810 \\
mean     & 0.0451 & 0.0095 & 0.0444 & 2.780 \\
min--max & 0.0436--0.0478 & 0.0091--0.0103 & 0.0437--0.0454 & 2.724--2.810 \\
\bottomrule
\end{tabular}
\end{table}

\begin{table}[h]
\centering\small
\caption{The same initialization sweep for the \emph{from-scratch control} (random initialization, no
pretrained backbone), at the same allocation, budget and data draw. The $\Cl$ range is $4.1\times$
(NLF-Turb) and $2.1\times$ (NLF-Trans) wider than the pretrained range of Table~\ref{tab:seedpre}.}
\label{tab:seedscratch}
\vspace{2pt}
\setlength{\tabcolsep}{6pt}
\begin{tabular}{lcccc}
\toprule
& $\Cl$ MAE & $\Cm$ MAE & $C_p$ surface MAE & Field rel-$L_2$ [\%] \\
\midrule
\multicolumn{5}{@{}l}{\textit{NLF-Turb}} \\
init 1   & 0.0537 & 0.0150 & 0.0673 & 2.319 \\
init 2   & 0.0676 & 0.0175 & 0.0748 & 2.440 \\
init 3   & 0.0582 & 0.0155 & 0.0710 & 2.327 \\
mean     & 0.0598 & 0.0160 & 0.0710 & 2.362 \\
min--max & 0.0537--0.0676 & 0.0150--0.0175 & 0.0673--0.0748 & 2.319--2.440 \\
\midrule
\multicolumn{5}{@{}l}{\textit{NLF-Trans}} \\
init 1   & 0.0605 & 0.0131 & 0.0633 & 3.418 \\
init 2   & 0.0640 & 0.0130 & 0.0635 & 3.429 \\
init 3   & 0.0693 & 0.0151 & 0.0665 & 3.513 \\
mean     & 0.0646 & 0.0137 & 0.0644 & 3.453 \\
min--max & 0.0605--0.0693 & 0.0130--0.0151 & 0.0633--0.0665 & 3.418--3.513 \\
\bottomrule
\end{tabular}
\end{table}

\section{Dependence of the gain on the fine-tuning budget}
\label{app:decay}

\begin{figure}[h]
\centering
\includegraphics{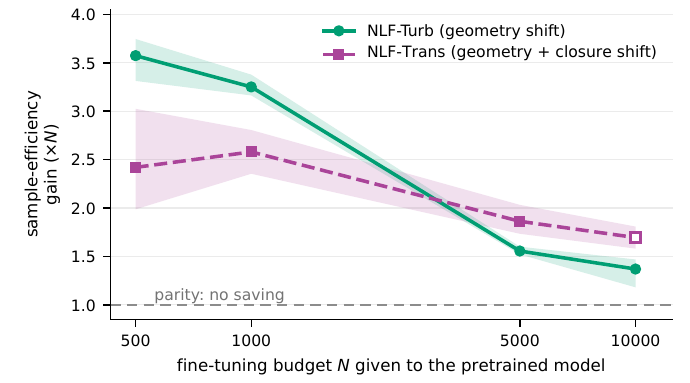}
\caption{Sample-efficiency gain on lift against fine-tuning budget $N$, both targets, under the
matched-draw estimator of \appref{app:uncertainty}: points are means over the three data draws and
shaded ribbons span them. At $N=10{,}000$ on NLF-Trans one data draw is censored --- its from-scratch
curve never reaches the pretrained error within the measured budgets --- so that point (open
marker) averages the remaining two. The curves cross between $N=1000$ and $N=5000$; the crossing is
not resolved within the measured budgets, so no location is marked.}
\label{fig:multcurve}
\end{figure}

Table~\ref{tab:headline} gives the sample-efficiency gains at the budgets the body quotes;
Fig.~\ref{fig:multcurve} plots them together with $N=500$. The geometry-shift target starts higher and decays faster, so
the curves cross; the crossing is bracketed by the measured budgets $N=1000$ and $N=5000$ and is not
resolved within them, so we rely on its existence rather than its location. The gain at budget $N$ is defined by the budget at which the from-scratch curve reaches the
pretrained model's error at $N$. At $N=20{,}000$ on NLF-Trans the pretrained error ($0.0297$) lies
below the from-scratch error at every measured budget including full data ($0.0300$), so the
read-off has no crossing within the data: the gain there is not zero but unmeasurable without
extrapolating the from-scratch curve. The plotted curve therefore ends at $N=10{,}000$, the last
budget at which the gain is defined. The same mechanism already censors one of the three draws at
$N=10{,}000$ (Table~\ref{tab:headline}, Fig.~\ref{fig:multcurve}).

\begin{table}[t]
\centering\footnotesize
\caption{Held-out lift error and sample-efficiency gain, depth allocation. Errors are means
over three data draws; parenthesized ranges are the per-draw spread. \textbf{Bold} marks the gains at the budget
the body reports, under the matched-draw estimator (\appref{app:uncertainty}). Full data is
$36{,}412$ and $34{,}585$ samples, run once. $^{\dagger}$For one of the three data draws the
from-scratch curve remains above the pretrained model's error at every measured budget, so no
matching budget exists and that draw's gain is undefined (censored); the reported mean averages the
two measurable draws.}
\label{tab:headline}
\vspace{2pt}
\setlength{\tabcolsep}{4.5pt}
\begin{tabular}{lccccccc}
\toprule
& \multicolumn{3}{c}{NLF-Turb (geometry shift)} & & \multicolumn{3}{c}{NLF-Trans (geometry $+$ transition modeling)} \\
\cmidrule(lr){2-4}\cmidrule(lr){6-8}
Budget & pretrained & scratch & gain & & pretrained & scratch & gain \\
\midrule
%% Per-draw gains at N=1000 verified against the corrected estimator (per-seed values
%% 3.163/3.212/3.378 and 2.354/2.579/2.806); body Q2 quotes the same ranges.
$N=1000$     & 0.0377 & 0.0529 & \textbf{3.25$\times$} (3.16--3.38) & & 0.0457 & 0.0571 & \textbf{2.58$\times$} (2.35--2.81) \\
% Gains are quoted to 2 d.p. throughout so the printed mean is never equal to an endpoint of
% its own range, and so the headline gap (3.25 - 2.58 = 0.67) reproduces from the table.

$N=5000$     & ---    & ---    & 1.56$\times$ (1.52--1.60)        & & ---    & ---    & 1.86$\times$ (1.73--2.03) \\
$N=10{,}000$ & ---    & ---    & 1.37$\times$ (1.18--1.47)          & & ---    & ---    & 1.70$\times^{\dagger}$ (1.58, 1.81) \\
full data    & 0.0262 & 0.0266 & ---                             & & 0.0287 & 0.0300 & --- \\
\bottomrule
\end{tabular}
\end{table}

\end{document}